\documentclass[article,1p,11pt]{elsarticle}
\usepackage[linktocpage=true]{hyperref}
\hypersetup{
colorlinks=true,
citecolor=MyDarkBlue,
linkcolor=MyDarkBlue,
urlcolor=MyDarkBlue
}
\usepackage[enableskew]{youngtab}
\usepackage{amsfonts,amsthm,amsmath,amssymb,natbib,color}
\usepackage[usenames,dvipsnames]{xcolor}
\definecolor{Red}{rgb}{1,0,0}
\def\Bbb{\mathbb}
\def\BZ{\mbox{$\Bbb Z$}} 

\def\BC{\mbox{$\Bbb C$}} 
\def\BP{\mbox{$\Bbb P$}}

\def\Bi{{\cal B}}
\def\Ba{{\cal B}_1}
\def\Bb{{\cal B}_2}
\def\Bc{{\cal B}_3}
\def\Bd{{\cal B}_4}
\def\Pc{{\BP}^3}
\newcommand{\sid}{\begin{equation}}
\newcommand{\sidd}{\end{equation}}

\makeatletter
\@addtoreset{equation}{section}
\makeatother

\journal{Nuclear Physics B}

\begin{document}

\begin{frontmatter}


\title{Partial resolution of complex cones over Fano $\Bi$}
\author{Siddharth Dwivedi}
\ead{siddharth@phy.iitb.ac.in}
\author{P. Ramadevi}
\ead{ramadevi@phy.iitb.ac.in}
\address{Department of Physics, Indian Institute of Technology Bombay,\\
 Mumbai, India, 400076}

\begin{abstract}
In our recent paper arXiv:1108.2387, we systematized inverse
algorithm to obtain quiver 
gauge theory living on the M2-branes probing the singularities of 
special kind of Calabi-Yau four-folds which were complex cones over 
toric Fano $\Pc$, $\Ba$, $\Bb$, $\Bc$.  These quiver gauge theories cannot 
be given a dimer tiling presentation. We use the method of partial resolution 
to show that the toric data of  ${\BC}^4$ and  Fano $\Pc$  can be 
embedded inside the toric data of Fano $\Bi$ theories. This method indirectly justifies that the two node quiver
Chern-Simons theories corresponding to ${\BC}^4$, Fano $\Pc$ and their orbifolds can be obtained by higgsing matter fields of the three node parent quiver corresponding to Fano $\Ba, \Bb,\Bc,\Bd$ three-folds.
\end{abstract}

\end{frontmatter}

\tableofcontents
\section{Introduction}
Initial works of Bagger-Lambert\cite{bag1,bag2,bag3} followed by
Gustavsson\cite{gus1,gus2}, Raamsdonk\cite{ram} and 
Aharony-Bergman-Jafferis-Maldacena (ABJM)\cite{abjm} led to a flurry
of interesting papers during the last four years  
between supersymmetric Chern-Simons  
gauge theory on  coincident $M2$-branes at the tip of Calabi-Yau four folds  
and their string duals. In a review article\cite{kleb}, these
developments are discussed in detail. 

Martelli et al\cite {mar} discussed the gauge-gravity correspondence
($AdS_4/CFT_3$) for some supersymmetric Chern-Simons theories with a 
quiver diagram description. Earlier works of Hanany et al in the context of Calabi-Yau three-folds\cite{han1} called \emph{forward 
algorithm}, can  be extended to obtain Calabi-Yau four-fold toric data
from $2+1$ dimensional quiver supersymmetric Chern-Simons theories. 

An elegant combinatorial approach called dimer tilings\cite{ken,he1}
which gives both the toric data and the corresponding quiver gauge theories was
generalised to study quiver Chern-Simons theories\cite{han3,yama,han4,han5,han6,han7a,han7b,han8}.
However, the dimer tiling approach is applicable for only a class of quiver gauge theories 
with $m$-matter fields, $r$ gauge group nodes and $N_W$ number
of terms in the superpotential $W$ satisfying $r-m+N_W=0$. The Chern-Simons (CS) levels
of the $r$-nodes can be denoted by the vector $\vec k= (k_1,k_2, \ldots k_r)$. 

One of the challenging problems was to determine 
quiver gauge theories corresponding to 18 toric Fano three-folds. A Fano variety in $d$-complex dimension is characterized by positive curvature and one can construct the CY ($d+1$)-fold by taking a complex cone over it. If the Fano variety is toric, the Calabi-Yau constructed from it will also be toric and one can attempt to find the dual quiver gauge theories. In 2 complex dimensions, there are 5 toric Fano 2-folds which are: zeroth Hirzebruch surface $\mathbb{F}_0$ and the del Pezzo surfaces $dP_0$, $dP_1$, $dP_2$, $dP_3$. The quiver gauge theories living on D3 branes probing the toric CY 3-folds obtained from these 5 Fano 2-folds have been studied in the literature\cite{han1}. Moving on to 3 complex dimensions, there are 18 toric Fano 3-folds\cite{wat,bat} with nomenclature (as used in \cite{han8}) Fanos $\mathbb{P}^3$, ${\cal{B}}_1$, ${\cal{B}}_2$, ${\cal{B}}_3$, ${\cal{B}}_4$, ${\cal{C}}_1$, ${\cal{C}}_2$, ${\cal{C}}_3$, ${\cal{C}}_4$, ${\cal{C}}_5$, ${\cal{D}}_1$, ${\cal{D}}_2$, ${\cal{E}}_1$, ${\cal{E}}_2$, ${\cal{E}}_3$, ${\cal{E}}_4$, ${\cal{F}}_1$, ${\cal{F}}_2$. The next problem was to obtain the quiver CS theory living on the M2 branes probing the toric CY 4-folds obtained from these 18 Fano 3-folds.

From the forward algorithm
and dimer tiling method, quiver gauge theories corresponding to fourteen of 
the toric Fanos ${\cal{B}}_4$, ${\cal{C}}_1$, ${\cal{C}}_2$, ${\cal{C}}_3$, ${\cal{C}}_4$, ${\cal{C}}_5$, ${\cal{D}}_1$, ${\cal{D}}_2$, ${\cal{E}}_1$, ${\cal{E}}_2$, ${\cal{E}}_3$, ${\cal{E}}_4$, ${\cal{F}}_1$, ${\cal{F}}_2$ were determined\cite{han8}. In Ref.\cite{sid},
we attempted the inverse algorithm of obtaining the quiver gauge theories
for the remaining four Fano 3-folds $\mathbb{P}^3$, ${\cal{B}}_1$, ${\cal{B}}_2$, ${\cal{B}}_3$. As expected, these quiver gauge theories do not satisfy $r-m+N_W=0$ confirming that they cannot be given dimer tiling presentation.

The next immediate question is to understand the embeddings inside the toric 
Fano  $\Bi$ three-folds by the method of partial resolution. In particular, we would like to 
obtain the quiver Chern-Simons theory corresponding to Fano ${\BP}^3$ by partial
resolution of Fano ${\cal B}$ three-folds.

Alternatively, we could determine embeddings 
from higgsing approach\cite{tapo,han7b,tapp}: 
For the quivers with $r$ nodes (usually called parent theories) which admit dimer tiling, we could higgsize some matter fields and obtain quivers with $r-1$ nodes (called daughter theories).  
Suppose we give a vacuum expectation value (VEV) to a bifundamental matter field $X_{ab}$ where the subscript denotes that the 
matter field is charged $+1$ with respect to node $a$ with Chern-Simons level $k_a$ and charged $-1$ 
with respect to node $b$ with Chern-Simons level $k_b$. This results in  coalescing of
two nodes $a$ and $b$ into a node with Chern-Simons level $k_a+k_b$.  
Higgsing of the three node quiver corresponding to a toric Fano ${\cal B}_4$
was discussed from dimer tilings in Ref.\cite{han7b}. In fact,  the daughter theories 
corresponds to either phase II of ${\BC}^4$ or phase I of ${\BC}^2/{\BZ}_2 \times {\BC}^2$ depending upon
which bifundamental field was given VEV.  

Another approach of higgsing called the algebraic method\cite{tapo} has been applied on parent 
quivers\cite{tapp} corresponding to some toric Fano 3-folds. We shall present the algebraic higgsing 
for Fano ${\cal B}_4$ and show that  the toric data  corresponding to daughter theories 
has to be ${\BC}^4$ or orbifolds of ${\BC}^4$. We will also study the method of partial 
resolution\cite{han1} and obtain the ${\BC}^4$ toric data for the daughter theories.

For the $3$-node quivers corresponding to  Fano ${\cal B}_1, {\cal B}_2$ and ${\cal B}_3$,
which do not admit dimer tiling presentation, we have to obtain daughter theories by the 
algebraic higgsing method\cite{tapp} or the partial resolution method\cite{han1}. We shall
show that the algebraic higgsing on these $3$-node quivers always gives ${\BC}^4$ or orbifolds of ${\BC}^4$. We also study the method of partial resolution to check whether the approach gives more information about the embedded theories. Interestingly this method shows that the toric data of Fano ${\BP}^3$ or its orbifolds is embedded in the
toric data of Fano ${\cal B}_2$ and Fano ${\cal B}_3$. 

The plan of the paper is as follows: In section \ref{sec2}, we briefly review the well known 2-node quiver Chern-Simons theory with four matter fields and
their toric data. In section \ref{sec3}, we will first perform the higgsing of toric Fano $\Bd$ using the algebraic method and determine the 
daughter quiver theories with $2$-nodes. Finally, we  show that the partial resolution method gives ${\BC}^4$ toric data.
In section \ref{sec4}, we will briefly present  the necessary data
of  the quiver corresponding to Fano $\Bc$. Then we
 study the algebraic higgsing and the method
of partial resolution for Fano ${\cal B}_3$.
Particularly, we show that the algebraic higgsing
gives ${\BC}^4$ or orbifolds of ${\BC}^4$ whereas the method
of partial resolution gives non-trivial
embeddings inside Fano ${\cal B}_3$.
In section \ref{sec5}, we study the method of partial
resolution for Fano ${\cal B}_2$.
We present the results of partial resolution of 
Fano $\Ba$ in section \ref{sec6} .
We summarize and discuss some open problems in section \ref{sec7}.
\section{Two-node quiver Chern-Simons theories}\label{sec2}
\begin{figure}
\centerline{\includegraphics[width=4in]{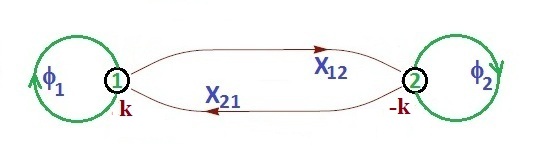}}
\caption{Quiver diagram (c)}
\label{fig:P3}
\end{figure}

Our aim is to study partial resolution of toric data corresponding to $3$-node parent
quiver resulting in a toric data corresponding to a
$2$-node daughter quiver. So,
in this section we will briefly recapitulate the $2$-node quivers with four
matter fields corresponding to toric data of
${\BC}^4$, orbifolds of ${\BC}^4$ and Fano ${\BP}^3$.
There can be three possible quivers:
\begin{enumerate}
\item Theory with 4 bi-fundamental matter fields $X_{12}^i$ and $X_{21}^i$
(where $i=1,2$), with CS levels ($k,-k$) and the superpotential  $W={\rm Tr}[\epsilon_{ij}X_{12}^1X_{21}^iX_{12}^2X_{21}^j]$. For the abelian groups, $W=0$. 
The quiver diagram is shown in figure~\ref{fig:Ga}. This theory admits tiling and the corresponding toric data is\cite{han7a}:
\begin{equation}
{\cal{G}}_a(k) = 
\left(
\begin{array}{cccc}
p_1 & p_2 & p_3 & p_4 \\
\hline 
 1 & 1 & 1 & 1 \\
 -1 & 0 & -1 & 0 \\
 0 & -1 & -1 & 0 \\
 0 & 0 & k & 0
\end{array}
\right)~.
\label{Ga}
\end{equation}
The charge matrix $Q_a$ for this theory given by $Q_a.{{\cal{G}}^T_a(k)} = 0$ is trivial, $Q_a = 0$. This theory is $\BZ_k$ orbifold of $\BC^4$, denoted as $\BC^4/\BZ_k$. For $k=1$, i.e, when the CS-levels of the theory is (1, -1), there is no orbifolding action and this theory in the literature is known as Phase-I of $\BC^4$\cite{han7a}.
\begin{figure}
\centerline{\includegraphics[width=4in]{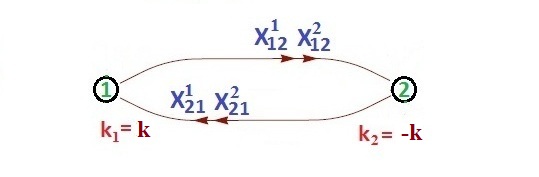}}
\caption{Quiver diagram (a)}
\label{fig:Ga}
\end{figure}
\item Theory with 2 adjoints $\phi_2^1, \phi_2^2$ present at the same node (say node 2) and two bifundamentals $X_{12},X_{21}$, with CS-levels ($k,-k$). Abelian $W$ of this theory is again zero
with non-abelian superpotential $W={\rm Tr }[X_{12}[\phi_2^1,\phi_2^2]X_{21}]$. The quiver diagram is shown in figure~\ref{fig:Gb}. This theory also admits tiling and the toric data for this theory is\cite{han7a}:
\begin{equation}
{\cal{G}}_b(k) = 
\left(
\begin{array}{cccc}
p_1 & p_2 & p_3 & p_4 \\
\hline 
 1 & 1 & 1 & 1 \\
 1 & 0 & 1 & 0 \\
 0 & -1 & 0 & 0 \\
 0 & 0 & k & 0
\end{array}
\right)
~.
\label{Gb}
\end{equation}
The charge matrix $Q_b$ of this theory given by $Q_b.{{\cal{G}}^T_b(k)}=0$ is also trivial, $Q_b = 0$. This is $\left(\BC^2/\BZ_k\right) \times \BC^2$ theory. For $k=1$, this theory is known as Phase-II of $\BC^4$.

It is pertinent to spell out the following obvious facts:\\
(i) We can deduce that the two quiver theories are distinct for $k \neq 1$ because the toric data of the two theories (${\cal{G}}_a(k)$ and ${\cal{G}}_b(k)$) are not related by any $GL(4, \BZ)$ transformation.\\ 
(ii) When there is no orbifolding- i.e., $k=1$, both the theories are same upto some $GL(4, \BZ)$ transformation. Hence these theories are actually just the phases of $\BC^4$ theory, known as Phase-I and Phase-II of $\BC^4$ respectively.
\begin{figure}
\centerline{\includegraphics[width=4in]{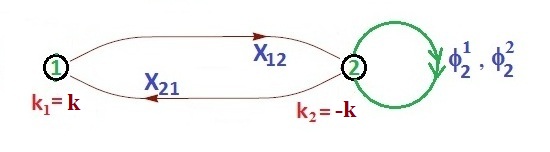}}
\caption{Quiver diagram (b)}
\label{fig:Gb}
\end{figure}
\item Theory with 2 bifundamentals $X_{12}$, $X_{21}$ and adjoints $\phi_1$, $\phi_2$ at two different nodes with trivial $W=0$. The quiver of this theory is shown in figure~\ref{fig:P3}. This theory does not admit tiling and was first obtained using the inverse algorithm in\cite{sid}, which was identified as the Fano $\mathbb{P}^3$ theory. 

However, we cannot determine the CS levels from the inverse algorithm\cite{sid}
for all the three $2$-node quiver theories. From the tiling description, 
we can obtain the CS levels for the
quiver diagrams (a) and (b).
Comparing the inverse algorithm of the three theories,    
we had inferred that the theory corresponding to Fano ${\BP}^3$ could 
have CS-levels $(4, -4)$.  As we do not know any method of determining the actual CS levels, the theory shown in quiver diagram (c) with $k=1$ could be Fano ${\BP}^3$ theory. 

A possible choice of toric data for the theory shown in figure \ref{fig:P3} can be:
\begin{equation}
{\cal{G}}_c(k) = 
\left(
\begin{array}{ccccc}
p_1 & p_2 & p_3 & p_4 & p_5 \\
\hline 
 1 & 1 & 1 & 1 & 1 \\
 1 & -1 & 0 & 0 & 0 \\
 0 & 1 & -1 & 0 & 0 \\
 0 & 0 & k & -k & 0
\end{array}
\right)
~,
\label{GP}
\end{equation}
which is related to the toric data  ${\cal G}_c(k=1)\equiv{\cal{G}}_{\BP^3}$ by $GL(4,{\BZ})$ transformation ${\cal T}(k)$, such that the determinant of $\cal T$ is $k$.   
\end{enumerate}
The charge matrix $Q_c(k)$  of this theory given by $Q_c(k).{{\cal{G}}_c(k)}^T=0$ consists only of $Q_F$ and is given by\cite{sid}:
\sid
Q_c(k) = Q_F^{{\BP }^3}= (1,1,1,1,-4)~, 
\label{QP3}
\sidd
for all $k$. This indicates that  the charge $Q_F$ will not  be able to
detect the orbifolding ${\BZ}_k$.

We would like to see these Calabi-Yau 4-folds, particularly 
Fano ${\BP}^3$, as embeddings inside Calabi-Yau 4-fold toric data corresponding to some $3$-node quivers. 
In the following section, we will extensively discuss the $3$-node quiver corresponding 
to Fano ${\cal B}_4$ which admits dimer tiling. 
Higgsing of the theory corresponding to Fano $\Bd$\cite{han7b}, from the tiling approach, has 
shown that the daughter theories are only quiver diagram (b). 
From the method of algebraic higgsing we get the toric data to be 
${\BC}^4$ and orbifolds  of ${\BC}^4$ whereas partial resolution gives toric data of $\BC^4$. This will set the necessary tools
and notations for studying the embeddings inside other three Fano ${\cal B}_i$'s in the later sections.  
\section{Fano $\Bd$} \label{sec3}
\begin{figure}
\centerline{\includegraphics[width=4in]{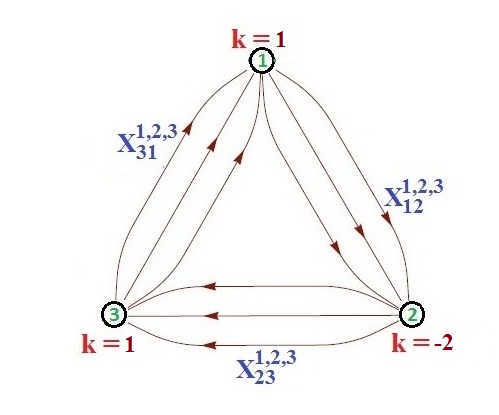}}
\caption{Quiver Diagram for Fano $\Bd$}
\label{fig:B4}
\end{figure}

The quiver corresponding to the complex cone over Fano ${\cal B}_4$ is 
a theory with $3$-nodes and $9$ bifundamental fields $X_{12}^i$, $X_{23}^i$, $X_{31}^i$, where $i=1,2,3$. This theory admits tiling and is known in the literature as Fano $\Bd$ or $M^{1,1,1}$ theory with CS-levels $\vec k=(k_1,k_2,k_3)=(1, -2, 1)$ \cite{han8}. 
The quiver diagram for this theory is shown in figure~\ref{fig:B4}.
The superpotential of the theory is given by:
\begin{equation}
W = {\rm Tr}\left[\epsilon_{ijk}X_{12}^i X_{23}^j X_{31}^k\right]
~.
\label{WB4}
\end{equation}
The incidence matrix of this theory is given by:
\begin{equation}
d = \left(
\begin{array}{c|ccccccccc}
& X_{12}^1 & X_{12}^2 & X_{12}^3 & X_{23}^1 & X_{23}^2 & X_{23}^3 & X_{31}^1 & X_{31}^2 & X_{31}^3 \\ \hline
G=1 & 1 & 1 & 1 & 0 & 0 & 0 & -1 & -1 & -1 \\
G=2 & -1 & -1 & -1 & 1 & 1 & 1 & 0 & 0 & 0 \\
G=3 & 0 & 0 & 0 & -1 & -1 & -1 & 1 & 1 & 1
\end{array}
\right)~,
\label{dB4}
\end{equation}
where the rows indicate the gauge groups or the nodes in the quiver, and columns indicate the matter fields.

The projected charge matrix ($\Delta$) will consist of a single row whose elements will be given by
\begin{equation}
\Delta_i = k_2d_{1i} - k_1d_{2i} = -(2d_{1i} + d_{2i})~.
\label{dtoD}
\end{equation}
Hence, $\Delta$ matrix will be given by:
\begin{equation}
\Delta = \left(
\begin{array}{ccccccccc}
X_{12}^1 & X_{12}^2 & X_{12}^3 & X_{23}^1 & X_{23}^2 & X_{23}^3 & X_{31}^1 & X_{31}^2 & X_{31}^3 \\ \hline
 -1 & -1 & -1 & -1 & -1 & -1 & 2 & 2  & 2 
\end{array}
\right)~.
\label{DB4}
\end{equation}
From the superpotential $W$ (\ref{WB4}), one can find the $F$-term constraints given by the set of equations $\left\{\partial W/\partial X_i = 0\right\}$, which means that the matter fields $X_i$'s can be written in terms of 5 independent $v$-fields, and the relation between them can be encoded in a matrix $K$. The matrix $K$ and its dual matrix $T$ are given as:
\sid
K=\left(\begin{array}{c|ccccc}
       {}&v_1&v_2&v_3&v_4&v_5\\
\hline
X_{12}^{(1)}&0& 0 & 1 & 0 & 0\\
X_{12}^{(2)}&0& 0 & 1 & 0 & 1\\
X_{12}^{(3)}&
0 & 0 & 1 & 1 & 0\\
X_{23}^{(1)}&
0 & 1 & 0 & 0 & 0\\
X_{23}^{(2)}&0
& 1 & 0 & 0 & 1\\
X_{23}^{(3)}&
0 & 1 & 0 & 1 & 0\\
X_{31}^{(1)}&
1 & 0 & 0 & 0 & 0\\
X_{31}^{(2)}&
1 & 0 & 0 & 0 & 1\\
X_{31}^{(3)}&
1 & 0 & 0 & 1 & 0\end{array}\right)~;~ T=\begin{pmatrix}1 & 0 & 0 & 0 & 0 & 1\cr 1 & 0 & 0 & 0 & 1 & 0\cr 
1 & 0 & 0 & 1 & 0 & 0\cr -1 & 0 & 1 & 0 & 0 & 0\cr -1 & 1 & 0 & 0 & 0 & 0\end{pmatrix}
~.
\label{kandt}
\sidd
From $K$ and $T$,  we can write the matrix $P = K.T$. The entries of 
the matrix $P$  are all non-negative and it gives the relation of the matter fields $X_i$'s with the GLSM fields $p_i$'s:
\sid
P=
\left(\begin{array}{c|cccccc}&p_1&p_2&p_3&p_4&p_5&p_6\\ \hline
X_{12}^1
&1& 0 & 0 & 1 & 0 & 0\\
X_{12}^2
&0 & 1 & 0 & 1 & 0 & 0\\ 
X_{12}^3
&0 & 0 & 1 & 1 & 0 & 0\\ 
X_{23}^1
&
1 & 0 & 0 & 0 & 1 & 0\\
X_{23}^2
&
0 & 1 & 0 & 0 & 1 & 0\\
X_{23}^3
&
0 & 0 & 
1 & 0 & 1 & 0\\
X_{31}^1
&1 & 0 & 0 & 0 & 0 & 1\\
X_{31}^2&0 & 1 & 0 & 0 & 0 & 1\\ 
X_{31}^3
&
0 & 0 & 1 & 0 & 0 & 1
\end{array}\right)~.
\label{p-mat}
\sidd
The $Q_F$ charge matrix is given by the nullspace of $P$:
\sid
Q_F = \left(
\begin{array}{cccccc}
1 & 1 & 1 & -1 & -1 & -1
\end{array}
\right)~.
\label{QFB4}
\sidd
The steps done so far in obtaining $P$-matrix and the charge $Q_F$ are independent of the CS levels. The information about CS levels is contained only in the charge $Q_D$ matrix which is a single row obeying the symmetry of the Calabi-Yau which is given as:  
\sid
Q_D = P. \Delta^t=\left[
\begin{array}{cccccc}
0 & 0 & 0 & 1 & 1 & -2
\end{array}
\right]~.
\label{QDB4}
\sidd
The total charge matrix can be obtained by concatenating (\ref{QFB4}) and (\ref{QDB4}) in a single matrix $Q$:
\sid
Q = \left(
\begin{array}{c}
Q_F \\ \hline
Q_D
\end{array}
\right) =
\left(
\begin{array}{cccccc}
1 & 1 & 1 & -1 & -1 & -1 \\ \hline
0 & 0 & 0 & 1 & 1 & -2
\end{array}
\right)~.
\label{QB4}
\sidd

The toric data of this theory is\cite{han8}:
\sid
{\cal{G}}(1, -2, 1) = \left(
\begin{array}{cccccc}
p_1 & p_2 & p_3 & p_4 & p_5 & p_6 \\ \hline
 1 & 1 & 1 & 1 & 1 & 1 \\
 1 & -1 & 0 & 0 & 0 & 0 \\
 0 & 1 & -1 & 0 & 0 & 0 \\
 0 & 0 & 0 & 1 & -1 & 0
\end{array}
\right)~.
\label{b4toric}
\sidd

Suppose we rescale the levels of this theory as $(k_1, k_2, k_3) =(n, -2n, n)$, where $n$ is some non-zero integer. Since this theory admits brane tiling description, the toric data for the scaled levels $(n, -2n, n)$ can be readily obtained from the Kastelyn matrix method\cite{han8}. The toric data is dependent on the scale $n$ and is given below:
\sid
{\cal{G}}(n, -2n, n) =
\left(
\begin{array}{cccccc}
 1 & 1 & 1 & 1 & 1 & 1 \\
 1 & 0 & -1 & 0 & 0 & 0 \\
 -1 & 1 & 0 & 0 & 0 & 0 \\
 0 & 0 & 0 & n & -n & 0
\end{array}
\right) 
\label{b4ntoric}
\sidd
Unfortunately,  when we perform forward algorithm for the 
scaled CS levels $(n,-2n,n)$ we see from eqns.(\ref {dtoD},\ref{QDB4}) that
$Q_D$ gets scaled by factor $n$ but $Q_F$ remains unchanged.
Clearly, 
the toric data ${\cal{G}}(1,-2,1)$  obtained as the null space of total charge matrix $Q$ i.e ${\cal{G}}.Q^t = 0$ is still satisfied by $Q^t=\left(\begin{matrix}Q_F^t
& n Q_D^t\end{matrix}\right)$. 

Equivalently, the toric data ${\cal G}$ obtained from forward 
algorithm has scaling ambiquity of any of the rows and hence
not unique. These arguments on ambiquity of $\cal G$, 
scaling of $Q_D$ and $\Delta$ under $\vec k \rightarrow n \vec k$ holds 
for inverse algorithm as well. 
Following the works on orbifolds of $\BC^4$, 
it is a well known fact that ${\cal{G}}(n, -2n, n)$ obtained from
tiling must be the $\mathbb{Z}_n$ orbifold of ${\cal{G}}(1, -2, 1)$ theory. 
The two toric datas are related by a $GL(4, \mathbb{Z})$ transformation ${\cal{T}}$ as:
\sid
{\cal{G}}(n, -2n, n) =  
{\cal{T}}.{\cal{G}}(1, -2, 1)~,
\sidd
where,
\sid
{\cal{T}} = 
 \left(
\begin{array}{cccc}
 1 & 0 & 0 & 0 \\
 0 & 1 & 0 & 0 \\
 0 & 0 & 1 & 0 \\
 0 & 0 & 0 & n
\end{array}
\right)~.
\sidd
Here $|det({\cal{T}})| =n$, which means that the volume of ${\cal{G}}(n, -2n, n)$ theory is $n$ times that of ${\cal{G}}(1, -2, 1)$ implying that the former theory is $\mathbb{Z}_n$ orbifold of latter theory. This  orbifolding
action, also scaling of CS levels,  is explainable only for those theories which admit tiling. From our forward algorithm discussion, we have explicitly
seen that the  scaling of CS levels does not give the orbifolding
information in the toric data.  Reconciling with tiling
results, we will hitherto use that fact that the scaling of CS levels 
represents ${\BZ}_n$ orbifolding of all theories which may
or may not admit tiling.

From the tiling approach, higgsing of the quiver diagram in figure~\ref{fig:B4} has been presented in
detail in Ref.\cite{han7b}. In particular, we can obtain daughter quivers in figure~\ref{fig:Gb} corresponding to ${\BC}^4$ and orbifolds of ${\BC}^4$. We will now study the algebraic approach of higgsing which will be applicable for other quivers that do not admit dimer tiling presentation. 

\subsection{Algebraic higgsing}
We  attempt the higgsing of Fano $\Bd$ theory to obtain one of the 2 nodes theories discussed in section \ref{sec2}. In algebraic higgsing, we choose some matter field, 
say $X_i$ and give a non zero VEV to it. Giving a VEV makes the matter field massive and hence removed from the quiver. However, in the process of giving VEV, all those GLSM $p_j$ fields which contain the matter field $X_i$ also become massive and hence must be removed. To do this, we delete the $i$-th row from the $P$ matrix which corresponds to the matter field $X_i$ and also, we remove all those columns ($p$-fields) which have non-zero entry corresponding to $i$-th row (and hence have become massive). 

As an example, let us take the $X_{12}^1$ field of the $\Bd$ theory and give a VEV to it. Thus, from the $P$ matrix (\ref{p-mat}), we must remove the first row and also the columns $1$ and 
$4$ which correspond to the GLSM fields  $p_1, p_4$ which contain the $X_{12}^1$ field. After giving VEV to the $X_{12}^1$ field, the nodes 1 and 2 in the figure~\ref{fig:B4} are collapsed giving a 2 node daughter theory with CS levels $(1,-1)$.
Removal of the corresponding row and columns in eqn.(\ref{p-mat}) will give the following reduced $P$-matrix: 
\sid
P_r =
\left(\begin{array}{c|cccc}
&p_2&p_3&p_5&p_6\\ \hline
X_{12}^2 & 1 & 0 & 0 & 0\\ 
X_{12}^3 & 0 & 1 & 0 & 0\\ 
X_{23}^1 & 0 & 0 & 1 & 0\\
X_{23}^2 & 1 & 0 & 1 & 0\\
X_{23}^3 & 0 & 1 & 1 & 0\\
X_{31}^1 & 0 & 0 & 0 & 1\\
X_{31}^2 & 1 & 0 & 0 & 1\\ 
X_{31}^3 & 0 & 1 & 0 & 1
\end{array}\right)~.
\sidd
The nullspace of this matrix gives the reduced charge matrix $Q_{F_r} = 0$. Thus, we see that the $Q_{F_r}$ of the daughter theory is trivial. The quivers (a) and (b) listed in section \ref{sec2}
have $Q_F=0$. So, algebraic higgsing can give daughter quivers (a) as well as quiver (b) with CS levels $(1,-1)$. 

By giving VEV to any other matter fields, we
have checked that we get the same trivial $Q_{F_r}=0$. This suggests that the algebraic higgsing of the parent quiver corresponding to Fano $\Bd$ will give quivers corresponding to ${\BC}^4$ and orbifolds of ${\BC}^4$.
It must be mentioned at this point that the algebraic higgsing of Fano $\Ba$, $\Bb$, $\Bc$ also gives a daughter theory with trivial $Q_F$. So, the algebraic higgsing tells that only $\BC^4$ or the orbifolds of $\BC^4$ are embedded inside Fano $\Ba,\Bb,\Bc,\Bd$. It may be possible that
we may get non-trivial $Q_F$ from the method
of partial resolution. So, we shall
study the method of partial resolution for
Fano $\Bd$ toric data.
\subsection{Partial Resolution}
In partial resolution, we try to remove the points from the toric diagram. The resulting toric diagram corresponds to some daughter theory which is embedded in the parent theory. From the toric diagram of $\Bd$, we will remove some points which amounts to removing the corresponding columns (or the corresponding $p$-fields) from the toric data ${\cal{G}}$. Next, we will check whether this reduced toric data (denoted as ${\cal{G}}_r$), which is obtained by removing the columns from original ${\cal{G}}$, is related to any of the toric data of the 2-node theories.

Take the toric data ${\cal{G}}$ of Fano $\Bd$
in eqn.(\ref{b4toric}).
We see that if we remove the points $p_1$ and $p_4$, we get the reduced toric data 
\sid
{\cal{G}}_r = \left(
\begin{array}{cccc}
 1 & 1 & 1 & 1 \\
 -1 & 0 & 0 & 0 \\
 1 & -1 & 0 & 0 \\
 0 & 0 & -1 & 0
\end{array}
\right)~.
\label{GrB4}
\sidd

This reduced toric data (\ref{GrB4}) is equivalent to the toric datas ${\cal{G}}_a(k=1)$ (\ref{Ga}) and ${\cal{G}}_b(k=1)$ (\ref{Gb}): 
\sid
{\cal{G}}_a(k=1) = \left(
\begin{array}{cccc}
 1 & 0 & 0 & 0 \\
 0 & 1 & 0 & 1 \\
 0 & 1 & 1 & 1 \\
 0 & 0 & 0 & -1
\end{array}
\right)~.~{\cal{G}}_r~,
\sidd
\sid
{\cal{G}}_b(k=1) =\left(
\begin{array}{cccc}
1 & 0 & 0 & 0 \\
0 & -1 & 0 & -1 \\
0 & 1 & 1 & 0 \\
0 & 0 & 0 & -1
\end{array}
\right)~.~{\cal{G}}_r~.
\sidd
Thus partial resolution of $\Bd$ toric data (\ref{b4toric}) only gives $\BC^4$ toric data. Explicit coalescing of nodes in tiling/algebraic higgsing allows both $\BC^4$ or $\BZ_2$ orbifold of $\BC^4$ whereas we get only ${\cal{G}}_a(k=1)$ or ${\cal{G}}_b(k=1)$ in the partial resolution method.

Alternatively, we can take
the charge matrix $Q$ of the $\Bd$ theory
given in eqn.(\ref{QB4}) and take a
linear combination and set the charge matrix elements corresponding
to column $p_1$ and $p_4$ to zero as described below:
Suppose that a row ($r$) of $Q$ of the daughter theory is given as some linear combination of rows ($R_i$) of $Q$ (\ref{QB4}) of parent theory, i.e.
\begin{equation}
r = \sum_{i=1}^2 (a_i R_i) = (a_1,a_1,a_1,-a_1+a_2,-a_1+a_2,-a_1-2a_2)~.
\end{equation}
Since we are removing $p_1$ and $p_4$ points, the corresponding columns in $r$ are set to 0 and also removed. This will give the values of $a_1$ and $a_2$, and we find that $a_1=a_2=0$. Hence, the $Q$ of the daughter theory is trivial.  

We have obtained ${\cal G}_r$ by
removing  all other possible set of points in the toric data of $\Bd$- namely., $\left\{p_1, p_5\right\}$, $\left\{p_2, p_4\right\}$, $\left\{p_2, p_5\right\}$, $\left\{p_3, p_4\right\}$ and $\left\{p_4, p_5\right\}$ 
and checked that they are again
related by $GL(4,\BZ)$ to ${\cal{G}}_a(k=1)$ and ${\cal{G}}_b(k=1)$. For all these cases, the
linear combination of the charge $Q$ matrix
(\ref {QB4}) with the appropriate
columns removed gives $Q$ of the daughter
theory to be trivial ($Q=0$). Thus we see that $\BC^4$ theory is embedded in the $\Bd$ theory. 

The quiver gauge theories corresponding to 
Fano $\Bc$ and $\Bb$ have non-trivial 
superpotential $W$\cite{sid} and hence can be described
by both forward and inverse algorithm. 
We will briefly present in the
next section, the 
quiver and necessary data for Fano 
$\Bc$ and study algebraic higgsing
and partial resolution.
\section{Fano $\Bc$} \label{sec4}
\begin{figure}
\centerline{\includegraphics[width=4in]{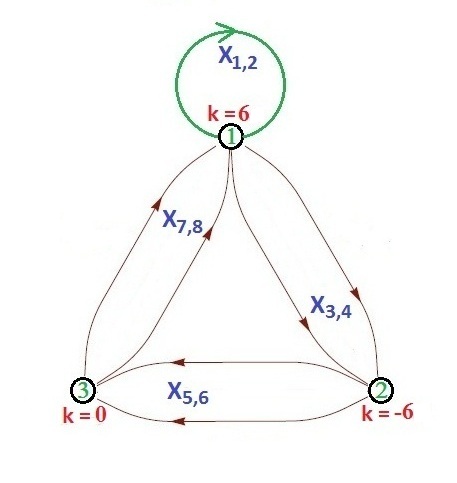}}
\caption{Quiver Diagram for Fano $\Bc$}
\label{fig:B3}
\end{figure}

It is a theory with 3-nodes, 2 adjoint fields $X_1, X_2$ on node-1 and 6 bifundamental fields $X_3, X_4, X_5, X_6, X_7, X_8$. The quiver diagram for this theory is shown in figure~\ref{fig:B3}. This theory does not admit tiling and was first studied in \cite{sid}, where it was identified as the quiver gauge theory for Fano $\Bc$ with CS-levels (6,-6,0).
The superpotential of the theory is given by:
\sid
W = {\rm Tr}\left[\left(X_1X_4-X_2X_3\right)\left(X_5X_8-X_6X_7\right)\right]~.
\label{WB3}
\sidd
From $W$ given in (\ref{WB3}), we construct $K$, which gives $T$ and hence $P$:
\sid
P = \left(
\begin{array}{c|ccccccc}
& p_1 & p_2 & p_3 & p_4 & p_5 & p_6 & p_7 \\ \hline
X_1 & 3 & 0 & 1 & 0 & 0 & 0 & 1 \\
X_2 & 0 & 3 & 1 & 0 & 0 & 0 & 1 \\
X_3 & 3 & 0 & 0 & 1 & 0 & 0 & 1 \\
X_4 & 0 & 3 & 0 & 1 & 0 & 0 & 1 \\
X_5 & 0 & 0 & 4 & 0 & 6 & 0 & 1 \\
X_6 & 0 & 0 & 0 & 4 & 6 & 0 & 1 \\
X_7 & 0 & 0 & 4 & 0 & 0 & 6 & 1 \\
X_8 & 0 & 0 & 0 & 4 & 0 & 6 & 1
\end{array}
\right)~.
\label{PB3}
\sidd
Total charge matrix $Q$ is given by:
\sid
Q = \left(
\begin{array}{c}
 Q_F \\ \hline
 Q_D
 \end{array}
\right)
=
\left(
\begin{array}{ccccccc}
 1 & 1 & 3 & 3 & -1 & -1 & -6 \\
 1 & 1 & 1 & 1 & 0 & 0 & -4 \\ \hline
0 & 0 & 2 & 2 & -2 & 0 & -2
\end{array}
\right)~.
\label{QB3}
\sidd
The toric data of the theory is given as \cite{sid}:
\sid
{\cal{G}}(6,-6,0) = \left(
\begin{array}{ccccccc}
p_1 & p_2 & p_3 & p_4 & p_5 & p_6 & p_7 \\ \hline
 1 & 1 & 1 & 1 & 1 & 1 & 1 \\
 1 & -1 & 0 & 0 & 0 & 0 & 0 \\
 0 & 0 & 1 & -1 & 0 & 0 & 0 \\
 0 & 1 & 0 & -1 & -1 & -1 & 0
\end{array}
\right)~.
\label{GB3}
\sidd
Similar to Fano $\Bd$ theory, if we rescale the levels here to $(k_1, k_2, k_3) =(6n, -6n, 0)$, where $n$ is some non-zero integer, we can write the toric data which has information about this scaling. Note that this theory does not admit tiling and the forward/inverse algorithm does not give an explicit dependence of ${\cal{G}}(6n, -6n, 0)$ on $n$, but there is a scaling ambiquity of any row of toric data. A choice of the
toric data will be 
\sid
{\cal{G}}(6n, -6n, 0) = 
\left(
\begin{array}{ccccccc}
 1 & 1 & 1 & 1 & 1 & 1 & 1 \\
 1 & -1 & 0 & 0 & 0 & 0 & 0 \\
 0 & 0 & 1 & -1 & 0 & 0 & 0 \\
 0 & n & 0 & -n & -n & -n & 0
\end{array}
\right)
\sidd
This toric data is related to that of ${\cal{G}}(6, -6, 0)$ by a volume factor $n$ and hence represents the $\mathbb{Z}_n$ orbifolding:
\sid
{\cal{G}}(6n, -6n, 0) = 
\left(
\begin{array}{cccc}
 1 & 0 & 0 & 0 \\
 0 & 1 & 0 & 0 \\
 0 & 0 & 1 & 0 \\
 0 & 0 & 0 & n
\end{array}
\right).{\cal{G}}(6, -6, 0)
\sidd 
We will now study algebraic
higgsing and obtain two-node daughter quivers.
\subsection{Algebraic higgsing}
If we give a VEV to any of the $X_i$ fields, thereby removing the corresponding row and columns from the $P$-matrix (\ref{PB3}), we see that we get a reduced matrix ($P_r$), whose null space (${Q_F}_r$) is always trivial. Starting from the parent quiver as shown in figure~\ref{fig:B3}, we see that the  non-trivial CS levels of the 2-node daughter quiver to be   $(6,-6)$. Thus, we can only say that the higgsing of the $\BZ_{n}$ orbifolds of Fano $\Bc$ theory will give $\BC^4/ \BZ_{6n}$ as the daughter theory. 
Moreover, we see that the last column ($p_7$) of P matrix (\ref{PB3}) which corresponds to the internal point in the toric diagram of Fano $\Bc$ will always be removed because it contains all the
matter fields. So, this method cannot give
a daughter theory corresponding to Fano ${\BP}^3$
which has an internal point in the toric diagram.
Hence we are forced to study the method
of partial resolution to check whether 
toric Fano ${\BP}^3$ is embedded inside $\Bc$ Fano.
\subsection{Partial Resolution}
Here, we do the partial resolution of Fano $\Bc$ theory and also check whether we get non-trivial
charge $Q$  for the daughter theory.
\subsubsection{Embedding of Fano $\BP^3$ inside Fano $\Bc$}
It is interesting to see that the
method of partial resolution does 
embed the toric Fano ${\BP}^3$ inside Fano $\Bc$
giving the correct $Q_F$ (\ref{QP3}). Hence we 
can claim that this method gives more information
than algebraic higgsing for quiver which do not
admit dimer tiling presentation.
Suppose we remove the points $\left\{p_5, p_6\right\}$ from the ${\cal{G}}$ (\ref{GB3}), we get a reduced toric data 
\sid
{\cal{G}}_r = \left(
\begin{array}{ccccc}
p_1 & p_2 & p_3 & p_4 & p_7 \\ \hline
 1 & 1 & 1 & 1 & 1 \\
 1 & -1 & 0 & 0 & 0 \\
 0 & 0 & 1 & -1 & 0 \\
 0 & 1 & 0 & -1 & 0
\end{array}
\right)
\sidd
and the toric data ${\cal{G}}_{\BP^3}\equiv{\cal G}_c(k=1)$ (\ref{GP}) is related to ${\cal{G}}_r$ by a $GL(4, \BZ)$ transformation:
\sid
{\cal{G}}_{\BP^3} = \left(
\begin{array}{cccc}
 1 & 0 & 0 & 0 \\
 0 & 1 & 0 & 0 \\
 0 & 0 & -1 & 1 \\
 0 & 0 & 1 & 0
\end{array}
\right).~{\cal{G}}_r~.
\sidd
A row ($r$) of $Q$ of the daughter theory will be given as linear combination of rows ($R_i$) of $Q$ (\ref{QB3}) i.e,
{\small
\begin{equation}
r = \sum_{i=1}^3 (a_i R_i) = (a_1+a_2, a_1+a_2, 3a_1+a_2+2a_3, 3a_1+a_2+2a_3,-a_1-2a_3,-a_1, -6a_1-4a_2-2a_3)~.
\nonumber
\end{equation}
Setting the columns 5,6 in $r$ to 0 gives $a_1 = a_3 = 0$. Removing these columns gives the reduced charge matrix as:
$$
r = (a_2, a_2, a_2, a_2, -4a_2) = a_2(1,1,1,1,-4)~.
$$
Thus, $Q$ will have only one row generated by $(1,1,1,1,-4)$. Hence, the charge matrix of 
daughter theory is same as charge matrix of the Fano ${\BP}^3$ (\ref{QP3}). Thus, we see that Fano $\BP^3$ is embedded in the Fano $\Bc$ theory. Similar to the partial resolution of $\Bd$ theory, we only get toric data of Fano $\BP^3$ but not the $\BZ_6$ orbifold of $\BP^3$, which is expected from the coalescing of the nodes in figure \ref{fig:B3}.
\subsubsection{Embedding of $\BC^4$ inside Fano $\Bc$}
If we remove points $\left\{p_1, p_3, p_5\right\}$ from the toric data (\ref{GB3}) we will get a reduced toric data given by: 
\sid
{\cal{G}}_r =\left(
\begin{array}{cccc}
p_2 & p_4 & p_6 & p_7 \\ \hline
 1 & 1 & 1 & 1 \\
 -1 & 0 & 0 & 0 \\
 0 & -1 & 0 & 0 \\
 1 & -1 & -1 & 0
\end{array}
\right)~.
\label{GrB3}
\sidd
Similar to the case of $\Bd$ theory, ${\cal{G}}_a(k=1)$ and ${\cal{G}}_b(k=1)$ are related to (\ref{GrB3}):
\sid
{\cal{G}}_a(k=1) = \left(
\begin{array}{cccc}
 1 & 0 & 0 & 0 \\
 0 & 2 & -1 & 1 \\
 0 & 1 & 0 & 1 \\
 0 & -1 & 1 & -1
\end{array}
\right).~{\cal{G}}_r~,
\sidd
\sid
{\cal{G}}_b(k=1) = \left(
\begin{array}{cccc}
 1 & 0 & 0 & 0 \\
 0 & -2 & 1 & -1 \\
 0 & 0 & 1 & 0 \\
 0 & -1 & 1 & -1
\end{array}
\right).~{\cal{G}}_r~.
\sidd
Also, we can find the reduced charge matrix $Q_r$ in a similar way as was done for other cases, and we find 
that it is trivial. Similarly, if we remove the other set of points $\left\{p_1, p_4, p_5\right\}$, 
$\left\{p_2, p_3, p_5\right\}$ and $\left\{p_2, p_4, p_5\right\}$, we will 
get trivial $Q_r$ and the reduced toric data in each case is only related to ${\cal{G}}_a$ and ${\cal{G}}_b$ toric datas for $k=1$ which implies $\BC^4$ toric data.

In the following section, we will briefly present the quiver corresponding to Fano $\Bb$ and study the partial resolution.
\section{Fano $\Bb$} \label{sec5}
The quiver Chern-Simons theory corresponding to Fano $\Bb$ \cite{sid} has 3 nodes, 12 matter fields with 
2 possible quiver diagrams as shown in figure~\ref{fig:B2-1} and figure~\ref{fig:B2-2} with CS levels $(2,-2,0)$
The superpotential of the theory is given by:
\begin{eqnarray}
W &=&{\rm Tr}\left( X_1X_4X_8X_{12} - X_1X_4X_9X_{11} - X_2X_5X_7X_{12}\right.\nonumber\\
~&~&\left. + X_2X_5X_9X_{10} + X_3X_6X_7X_{11} - X_3X_6X_8X_{10}\right)~.
\label{WB2}
\end{eqnarray}
This theory does not admit dimer tiling presentation but can be studied using 
forward algorithm. 
\begin{figure}
\centerline{\includegraphics[width=4in]{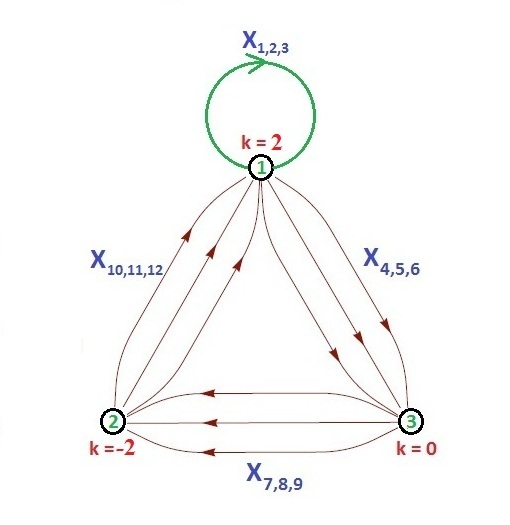}}
\caption{Cyclic quiver for Fano $\Bb$}
\label{fig:B2-1}
\end{figure}
\begin{figure}
\centerline{\includegraphics[width=4in]{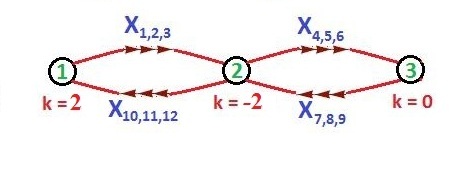}}
\caption{Linear quiver for Fano $\Bb$}
\label{fig:B2-2}
\end{figure}
Using $W$ given in eqn.(\ref{WB2}), one can construct $K$ which gives $T$ whence $P$:
\sid
P = \left(
\begin{array}{c|cccccccc}
& p_1 & p_2 & p_3 & p_4 & p_5 & p_6 & p_7 & p_8 \\ \hline
X_1 & 2 & 0 & 0 & 2 & 2 & 0 & 0 & 1 \\
X_2 & 0 & 2 & 0 & 2 & 2 & 0 & 0 & 1 \\
X_3 & 0 & 0 & 2 & 2 & 2 & 0 & 0 & 1 \\
X_4 & 2 & 0 & 0 & 0 & 2 & 2 & 0 & 1 \\
X_5 & 0 & 2 & 0 & 0 & 2 & 2 & 0 & 1 \\
X_6 & 0 & 0 & 2 & 0 & 2 & 2 & 0 & 1 \\
X_7 & 4 & 0 & 0 & 2 & 0 & 0 & 2 & 1 \\
X_8 & 0 & 4 & 0 & 2 & 0 & 0 & 2 & 1 \\
X_9 & 0 & 0 & 4 & 2 & 0 & 0 & 2 & 1 \\
X_{10} & 4 & 0 & 0 & 0 & 0 & 2 & 2 & 1 \\
X_{11} & 0 & 4 & 0 & 0 & 0 & 2 & 2 & 1 \\
X_{12} & 0 & 0 & 4 & 0 & 0 & 2 & 2 & 1
\end{array}
\right)~.
\label{PB2}
\sidd
A possible choice of the total charge matrix $Q$ is given by:
\sid
Q = \left(
\begin{array}{c}
 Q_F \\ \hline
 Q_D
\end{array}
\right)
=
\left(
\begin{array}{cccccccc}
 1 & 1 & 1 & -1 & -1 & -1 & -2 & 2 \\
 0 & 0 & 0 & 1 & -1 & 1 & -1 & 0 \\
 0 & 0 & 0 & 0 & 1 & 0 & 1 & -2 \\ \hline
 0 & 0 & 0 & 1 & 1 & 2 & 0 & -4
\end{array}
\right)~.
\label{QB2}
\sidd
The toric data of the theory is given as\cite{sid}:
\sid
{\cal{G}}(2,-2,0) = \left(
\begin{array}{cccccccc}
p_1 & p_2 & p_3 & p_4 & p_5 & p_6 & p_7 & p_8 \\ \hline
 1 & 1 & 1 & 1 & 1 & 1 & 1 & 1 \\
 1 & -1 & 0 & 0 & 0 & 0 & 0 & 0 \\
 0 & 1 & -1 & 0 & 0 & 0 & 0 & 0 \\
 0 & 0 & 1 & -1 & -1 & 1 & 1 & 0
\end{array}
\right)~.
\label{GB2}
\sidd
Taking the  $P$ matrix (\ref{PB2}), giving VEV to any of the matter fields 
gives ${Q_F}_r=0$. Also from coalescing of nodes in the quiver diagrams given in figure~\ref{fig:B2-1} and 
figure~\ref{fig:B2-2}, we know that the non-trivial CS level of the 2 node daughter theory will be $(2,-2)$. 
Thus, the daughter theory will be $\BZ_{2}$ orbifolds of $\BC^4$. 
From eqn.(\ref{PB2}) we see that giving a VEV to any of the matter fields will always remove the 
last column ($p_8$) of the $P$ which corresponds to an internal point in the toric diagram of $\Bb$. 
So if we do the algebraic higgsing of $\Bb$, we are never going to get the embedding as Fano $\BP^3$   
which has an internal point in its toric diagram. We will now work out the partial resolution to 
see if we can get more information. 
\subsection{Partial Resolution}
In this case, we found that the if we remove the set of points 
$\left\{p_1,p_4,p_5,p_7\right\}$, $\left\{p_1,p_4,p_6,p_7\right\}$, 
$\left\{p_2,p_4,p_5,p_7\right\}$, $\left\{p_2,p_4,p_6,p_7\right\}$, 
$\left\{p_3,p_4,p_5,p_7\right\}$, $\left\{p_3,p_4,p_6,p_7\right\}$ or 
$\left\{p_4,p_5,p_6,p_7\right\}$, we will get a reduced toric data which is related to ${\cal{G}}_a(k=1)$ or ${\cal{G}}_b(k=1)$ by a $GL(4,\BZ)$ transformation. Further the nullspace, 
i.e. the reduced charge matrix $Q_r$ is trivial. This implies that the toric data of
the daughter theory is
$\BC^4$. 
\subsubsection{Embedding of Fano $\BP^3$ inside Fano $\Bb$}
If we remove the points $\left\{p_5, p_6, p_7\right\}$ from the toric data ${\cal{G}}$ given in eqn.(\ref{GB2}), 
we will get the reduced toric data:
$${\cal{G}}_r = \left(
\begin{array}{ccccc}
p_1 & p_2 & p_3 & p_4 & p_8 \\ \hline
 1 & 1 & 1 & 1 & 1 \\
 1 & -1 & 0 & 0 & 0 \\
 0 & 1 & -1 & 0 & 0 \\
 0 & 0 & 1 & -1 & 0
\end{array}
\right)~,$$ which is exactly same as the toric data of $\BP^3$. Thus we see that $\BP^3$ is 
embedded inside $\Bb$.

Taking a row ($r$) of total charge $Q$ of the daughter theory as a linear combination of 
rows ($R_i$) of total charge matrix of $\Bb$ given in eqn.(\ref{QB2}):
$$
r = (a_1, a_1, a_1, -a_1+a_2+a_4,-a_1-a_2+a_3+a_4,-a_1+a_2+2a_4,-2a_1-a_2+a_3, 2a_1-2a_3-4a_4)~.
$$
Setting the columns 5,6,7 in $r$ to 0, we get $a_1 = a_2/3 = a_3/5 = -a_4$. The reduced charge matrix after 
removal of the columns 5,6,7 gives:
$$
r = (a_1, a_1, a_1, a_1, -4a_1) = a_1(1,1,1,1,-4)~.
$$
Thus, $Q$ will have only 1 row generated by $(1,1,1,1,-4)$. Hence, the charge matrix of daughter theory is 
$Q = (Q_F) = (1,1,1,1,-4)$ which is the charge matrix of $\BP^3$. Thus partial resolution only gives $\BP^3$. However from coalescing of nodes in figures \ref{fig:B2-1} and \ref{fig:B2-2}, we expect the daughter theory to be $\BZ_2$ orbifold of $\BP^3$.
\section{Fano $\Ba$} \label{sec6}
This theory was studied in \cite{sid} where the inverse algorithm was used to find a quiver gauge theory 
shown in figure~\ref{fig:B1} with CS levels ($2,0,-2$). The superpotential of this theory was obtained as:
\sid
W = \text{Tr}\left[X_2X_5X_8\left(X_1X_4X_9X_3X_6X_7-X_1X_6X_7X_3X_4X_9\right)\right]~.
\label{WB1}
\sidd
Note that the abelian $W=0$ and the quiver does not admit tiling. So, there is no way to obtain the toric data 
of this theory using forward algorithm or dimer tiling approach. 
The toric data for $\Ba$ theory with multiplicity is given as \cite{sid}:
\sid
{\cal{G}}=
\left(
\begin{array}{ccccccc}
p_1 & p_2 & p_3 & p_4 & p_5 & p_6 & p_7 \\ \hline
 1 & 1 & 1 & 1 & 1 & 1 & 1 \\
 1 & -1 & 0 & 0 & 0 & 0 & 0 \\
 0 & 1 & -1 & 0 & 0 & 0 & 0 \\
 0 & 0 & 2 & -1 & 1 & 1 & 0
\end{array}
\right)~.
\label{GB1}
\sidd
From the ansatz given in \cite{sid}, we can take a choice of $Q_F$ and $Q_D$ obeying the symmetry of Calabi-Yau over Fano $\Ba$ and the total charge matrix will be given by:
\sid
Q = \left(
\begin{array}{c}
 Q_F \\ \hline
 Q_D
\end{array}
\right) = \left(
\begin{array}{ccccccc}
 1 & 1 & 1 & -2 & -2 & -2 & 3 \\
 0 & 0 & 0 & 2 & 1 & 1 & -4 \\ \hline
 0 & 0 & 0 & 1 & 0 & 1 & -2
\end{array}
\right)~.
\sidd

The algebraic higgsing in this case gives $\BZ_{2}$ orbifolds of $\BC^4$
and we also observe that the internal point in the toric Fano $\Ba$ gets removed. The partial resolution by 
removing the points $\left\{p_1,p_4,p_5\right\}$, $\left\{p_2,p_4,p_5\right\}$ or 
$\left\{p_3,p_4,p_5\right\}$ gives toric data of $\BC^4$. We could not obtain $\BP^3$ as embedding inside $\Ba$ by this method.
\begin{figure}
\centerline{\includegraphics[width=4in]{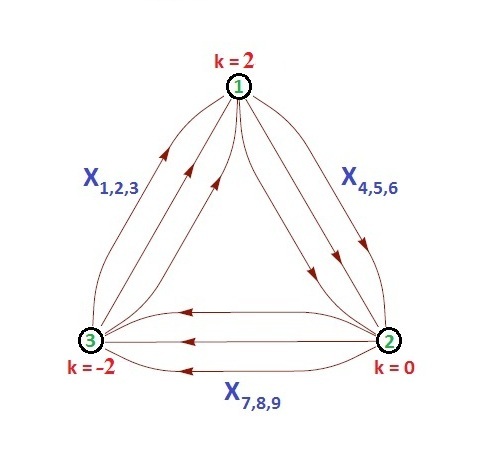}}
\caption{Quiver Diagram for Fano $\Ba$}
\label{fig:B1}
\end{figure}

\section{Conclusions}\label{sec7}
Our main motivation was to determine the $2$-node daughter quivers by the method of higgsing the 
$3$-node parent quivers.
Particularly, we wanted to higgsise matter fields
of the $3$-node quivers corresponding to Fano $\Ba,\Bb,\Bc,\Bd$
3-folds and obtain the daughter quivers. This procedure will determine the CS levels
of the daughter quiver theories. Unfortunately, the algebraic method of higgsing does
not give any non-trivial charge matrix $Q$.

 
For the quiver corresponding to Fano $\Bd$, which has
dimer tiling presentation, higgsing could be studied
by tiling approach as well as by algebraic method.
By the algebraic method of higgsing, we get the
reduced charge matrix $Q_r=0$ which suggests
that the two-node daughter theories can be
${\BC}^4$ and orbifolds of ${\BC}^4$. From the method of partial resolution of
the toric data corresponding to Fano $\Bd$, we obtained only $\BC^4$. 
We have shown that the scaling of CS levels $(1,-2,1) \rightarrow (n,-2n,n)$ 
results in ${\BZ}_n$  orbifolding of Fano $\Bd$ theory. Hence , partial resolution
of the orbifolded Fano $\Bd$ will give  $\BC^4 /\BZ_n$  toric data with trivial $Q_F$.  

The $3$-node quivers for other Fano $\Bi$ which were determined from the inverse algorithm do not 
admit dimer tiling presentation. So, we cannot study higgsing of these theories from tiling
approach. The algebraic higgsing of any matter field removes the
information of the internal point in the toric data and always
gives trivial reduced charge matrix $Q_r=0$. Unlike
${\BC}^4$ and its orbifolds, the toric data of
Fano ${\BP}^3$ has an internal point.   
So, we studied the method of partial resolution for toric Fano $\Ba,\Bb,\Bc$ 3-folds. 
We found that the Fano ${\BP}^3$ can be embedded inside Fano $\Bc$ and $\Bb$ theories. 

Algebraic higgsing and unhiggsing of quiver theories corresponding to some Fano $3$-folds
have been studied recently\cite{tapp}. Higgsing certain 
matter fields in the $4$-node quivers corresponding
to toric Fano ${\cal C}_4$ gives the $3$-node quiver corresponding to Fano $\Bd$.
Also higgsing of the quiver corresponding to Fano ${\cal D}_2$ gives
daughter quiver corresponding to Fano $\Bd$. 
These results can also be reproduced using the method of 
partial resolution. 

It is not obvious whether we can obtain other Fano $\Bi$'s toric data
as embeddings inside Fano ${\cal C}_4$ and Fano ${\cal D}_2$
$3$-folds. One of the issues is about our
choice of multiplicity of certain points in the toric data for 
Fano $\Ba,\Bb,\Bc$. 

We had chosen a charge matrix $Q_F,Q_D$ 
respecting the ansatz\cite{sid} which determined the multiplicity
of certain points in the toric data of Fano $\Bi$'s.
The toric data with the specific multiplicity of certain
points was important to obtain sensible quivers. 
In principle, we  would like to do the method of partial resolution for toric
data corresponding to  some $4$-node quivers and reproduce the
toric data of Fano $\Ba,\Bb,\Bc$ with the correct
multiplicity of some of the points. We hope to study in future the 
embeddings of 
the toric Fano $\Bi$'s inside toric four-folds corresponding to 
$4$-node quiver Chern-Simons theories. 

\vskip.5cm
\noindent
{\bf Acknowledgments}: We would like to thank A.Hanany
for discussions.  We are grateful to T.Sarkar and 
Rak-Kyeong for their valuable inputs on the orbifolding issues.


\vskip.5cm
\newpage

\begin{thebibliography}{99}
\bibitem{bag1}J.Bagger and N.Lambert, ``Modeling multiple M2's,'' 
Phys. Rev. {\bf D75}, 045020 (2007) [arXiv:hep-th/0611108].
\bibitem{bag2}J.Bagger and N.Lambert, ``Gauge symmetry and Supersymmetry of Multiple M2-Branes,'' Phys. Rev. {\bf D77},065008 
(2008) [arXiv:0711.0955[hep-th]].
\bibitem{bag3}J.Bagger and N.Lambert, ``Comments on Multiple M2-branes,'' JHEP{\bf 0802}, 105
(2008) [arXiv:0712.3738[hep-th]].
\bibitem{gus1}A.Gustavsson, ``Algebraic structures on parallel M2-branes,''  Nucl.Phys. {\bf B811}, 66 (2009)
[arXiv:0709.1260[hep-th]]. 
\bibitem{gus2}A.Gustavsson,``Selfdual strings and loop space Nahm equations,'' JHEP {\bf 0804}, 083 (2008)
[arXiv:0802.3456[hep-th]].
\bibitem{ram} M.Van Raamsdonk, ``Comments on the Bagger-Lambert theory and multiple
M2-branes,'' JHEP {\bf 0805}, 105 (2008) [arXiv:0803.3803].
\bibitem{abjm}O.Aharony, O. Bergman, D.L. Jafferis and J. Maldacena,
``N=6 superconformal Chern-Simons-matter theories, M2-branes and their gravity duals, ''JHEP {\bf 0810},091 (2008)
[arXiv:0806.1218[hep-th]].
\bibitem{kleb}I.R.Klebanov and G.Torri, ``M2-branes and AdS/CFT,'' Int.J.Mod.Phys. {\bf A25} 332 (2010) [arXiv:0909.1580[hep-th]].
\bibitem{mar}D.Martelli and J. Sparks, ``Moduli spaces of Chern-Simons
quiver gauge theories and AdS$_4$/CFT$_3$,'' Phys.Rev. {\bf D78}, 126005 (2008) [arXiv:0808.0912[hep-th]].
\bibitem{han1}B.Feng, A.Hanany and Y.H.He, ``D-brane gauge theories from toric singularities 
and toric duality,'' Nucl. Phys. {\bf B595}, 165 (2001) [arXiv:hep-th/0003085].
\bibitem{ken}A. Hanany and K.D. Kennaway, ``Dimer models and toric diagrams,''
[arXiv:hep-th/0503149].
\bibitem{he1}S.Franco, A.Hanany, K.D. Kennaway,D.Vegh and B.Wecht,
``Brane dimers and quiver gauge theories,'' JHEP {\bf 0601}, 096 (2006) [arXiv:hep-th/
0504110].
\bibitem{han3}A.Hanany and A.Zaffaroni, ``Tilings, Chern-Simons Theories and M2 branes,'' JHEP {\bf 0810}, 111 (2008)
[arXiv:0808.1244[hep-th]].
\bibitem{yama}K.Ueda and M.Yamazaki, ``Toric Calabi-Yau four-folds dual to Chern-Simons-matter
theories,'' JHEP {\bf 0812}, 045 (2008) [arXiv:0808.3768[hep-th]].
\bibitem{han4}A.Hanany, D.Vegh, A.Zaffaroni, ``Brane Tilings and M2 branes,'' JHEP {\bf 0903}, 012 (2009) [arXiv:0809.1440].
\bibitem{han5}S.Franco, A.Hanany, J.Park and D.Rodriguez-Gomez,``Towards M2-brane 
Theories for Generic Toric Singularities,''JHEP {\bf 0812}, 110 (2008) [arXiv:0809.3237
[hep-th]].
\bibitem{han6}A.Hanany and Y.H.He, ``M2-branes and Quiver Chern-Simons: A Taxonomic Study,''
arXiv:0811.4044[hep-th].
\bibitem{han7a}J.Davey, A.Hanany, N.Mekareeya and G.Torri, ``Phases of M2-brane Theories,'' JHEP {\bf 0906}, 025 (2009)
[arXiv:0903.3234[hep-th]].
\bibitem{han7b} J.Davey, A.Hanany, N.Mekareeya and G.Torri, ``Higgsing M2-brane Theories,''
arXiv:0908.4033[hep-th].
\bibitem{han8}J.Davey, A.Hanany, N.Mekareeya and G.Torri, ``M2-Branes and Fano 3-folds,''
arXiv:1103.0553[hep-th].
\bibitem{wat}K. Watanabe, M. Watanabe, Tokyo J. Math, \textbf{5}, no. 1 (1982)
\bibitem{bat}V. V. Batyrev, ``Toroidal Fano 3-folds,'' Math. USSR-Izv \textbf{19}, 13 (1982)
\bibitem{sid}S. Dwivedi, P. Ramadevi, ``Inverse algorithm and M2-brane theories,'' 
JHEP {\bf 1111}, 111 (2011)	[arXiv:1108.2387v3 [hep-th]].
\bibitem{tapo}P. Agarwal, P. Ramadevi, T. Sarkar, ``A note on dimer models and D-brane gauge 
theories,'' JHEP {\bf 0806} 054 (2008) [arXiv:0804.1902].
\bibitem{tapp}Prabwal Phukon, Tapobrata Sarkar, ``On the Higgsing and UnHiggsing of Fano 3-Folds,'' JHEP {\bf 1201} 090 (2012) [arXiv:1108.4237v1].
\end{thebibliography}
\end{document}